\documentstyle[13lomcon,cite,amssymb,epsfig]{article}

\def\ds{\displaystyle}

\begin{document}

\title{BILINEAR R-PARITY VIOLATION IN RARE MESON DECAYS}

\author{ A. Ali \footnote{e-mail: ahmed.ali@desy.de}}
\address{Deutsches Elektronen-Synchrotron, DESY, 22607 Hamburg, Germany}

\author{A. V. Borisov \footnote{e-mail: borisov@phys.msu.ru},
M. V. Sidorova \footnote{e-mail:  mvsid@rambler.ru}}

\address{Faculty of Physics, Moscow State University, 119991 Moscow, Russia}


\maketitle\abstracts{~We discuss rare meson decays $
K^ +   \to \pi ^ -  \ell ^ +  \ell '^ +$ and $D^ +   \to K^ -  \ell ^ +  \ell '^ +  $
($\ell ,\ell ' = e,\mu $) in a supersymmetric extension of the standard model with
 explicit breaking of $R$-parity by bilinear Yukawa couplings in the superpotential.
 Estimates of the branching ratios for these decays are given. We also compare our
 numerical results with analogous ones previously obtained for two other mechanisms
 of lepton number violation: exchange by massive Majorana neutrinos and trilinear
 $R$-parity violation.}


In the standard model (SM), the lepton $L$ and baryon $B$ numbers are conserved to all orders of perturbation theory due to the accidental $
U(1)_L  \times U(1)_B$ symmetry existing at the level of renormalizable operators. But the $L$ and $B$ nonconservation is a generic feature of various extensions of the SM \cite{moh}. That is why lepton-number (LN) violating processes have long been recognized as a sensitive tool to put theories beyond the SM to the test. One of the most well known process of such type is neutrinoless double beta decay $(A,Z) \to (A,Z + 2) + e^ -   + e^ -$  that has been searched for many years (see, e.g., \cite{abzh} and references therein).

In Refs. \cite{abz,abs1} rare decays of the pseudoscalar mesons $K$, $D$, $D_s$, and $B$ of the type
\begin{equation}
\label{dec}
M^ +   \to M'^ -  \ell ^ +  \ell '^ +  \;\left( {\ell ,\ell ' = e,\mu } \right)
\end{equation}
 mediated by light ($m_N  \ll m_\ell  , m_{\ell '} $ ) and heavy ($m_N  \gg m_M $) Majorana neutrinos were investigated. The indirect upper bounds on the branching ratios for the decays (\ref{dec}) have been derived taking into account the limits on lepton mixing and neutrino masses obtained from the precision electroweak measurements, neutrino oscillations, cosmological data, and searches of the neutrinoless double beta decay. These bounds are greatly more stringent than the direct experimental ones \cite{pdg}.

 In Refs. \cite{abs2,abs3} we considered another mechanism of the $\Delta L = 2$ decays (\ref{dec}) based on the  minimal supersymmetric extension of the SM with explicit  $R$-parity violation ($\not\!\!R$MSSM, for a review, see \cite{barb}). $R$-parity is defined as $R = ( - 1)^{3(B - L) + 2S} $, where $B$, $L$, and $S$ are the baryon and lepton numbers and the spin, respectively. The SM fields, including additional Higgs boson fields appearing in the extended models, have $R=1$ while $R= - 1$ for their superpartners.

The most general form for the $R$-parity and lepton number
violating part of the superpotential is given by \cite{barb}
\begin{equation}
\label{rpv}
W_{\not R}  = \varepsilon _{\alpha\beta } \left(
{\frac{1} {2}\lambda _{ijk} L_i^\alpha  L_j^\beta  \bar E_k  +
\lambda '_{ijk} L_i^\alpha  Q_j^\beta  \bar D_k  + \epsilon _i
L_i^\alpha H_u^\beta  } \right).
\end{equation}
Here $i, j, k =1, 2, 3$ are generation indices, $L$ and $Q$ are
$SU(2)$ doublets of left-handed lepton and quark superfields
($\alpha, \beta = 1, 2$ are isospinor indices), $\bar{E}$ and
$\bar{D}$ are singlets of right-handed superfields of leptons and
down quarks, respectively; $H_u$ is a doublet Higgs superfield
(with hypercharge $Y=1$); $\lambda _{ijk} (= - \lambda _{jik})
,~\lambda '_{ijk}$ and $\epsilon _i $ are trilinear and bilinear Yukawa couplings, respectively.

We assumed in \cite{abs2,abs3} that the bilinear couplings are absent at tree level ($\epsilon _i =0$ in Eq. (\ref{rpv})). As well known they are  generated by quantum corrections \cite{barb} but the dominant contribution of the tree-level trilinear couplings to the phenomenology is expected. In the present report, we investigate the case of tree-level bilinear couplings: $\epsilon _i \neq 0$ with $\lambda = 0$,  $\lambda' = 0$. For this case, trilinear couplings cannot be generated via radiative corrections.

The bilinear terms in the superpotential (\ref{rpv}) induce mixing between the SM leptons and the MSSM charginos and neutralinos $\tilde{\chi}_n^0$ in the mass-eigenstate basis and lead to the following $\Delta L = \pm 1$ lepton-quark operators \cite{fae,hir}:
\begin{eqnarray}
\label{lh}
& {\cal L}_{LH}= \ds
-\frac{g}{\sqrt{2}}\kappa_{n}W_{\mu}^-\bar{\ell}\gamma^{\mu}
P_L\tilde{\chi}_n^0
+\sqrt{2}g\left(\beta_k^d\bar{\nu}_kP_Rd\tilde{d}_R^*\right.\nonumber\\
&\left. + \beta_k^u\bar{\nu}_kP_Ru^c\tilde{u}_L+
\beta_{ki}^{\ell}\bar{\nu}_k
P_R\ell^c\tilde{\ell}_{Li}+\beta^c\bar{u}P_R\ell^c\tilde{d}_L
\right)+{\rm H.c.}
\end{eqnarray}
Here $P_{L,R}=(1\mp \gamma^5)/2$, $ \gamma^{5}
=i\gamma^{0}\gamma^{1}\gamma^{2}\gamma^{3}$; the constants $\kappa_{n}$ ($n=1,2,3,4$), $\beta_k^d$, $\beta_k^u$, $\beta_{ki}^{\ell}$ ($k,i=1,2,3$) and
$\beta^c$ depend on the elements of the mixing matrices diagonalizing the neutrino-neutralino and the charged lepton-chargino mass matrices.

The Lagrangian describing the bilinear mechanism of the decays (\ref{dec}) is
\begin{equation}
\label{bi}
{\cal L}= {\cal L}_{LH}+{\cal L}_{SM}+{\cal L}_{\tilde{g}}+{\cal L}_{\tilde{\chi}}.
\vspace{-0.2cm}
\end{equation}
In addition to the LN violating part (\ref{lh}), it includes the SM charged-current interactions
\begin{equation}
\label{SM}
{\cal L}_{SM}=\frac{g}{{\sqrt 2 }}W^{ + \mu } ( \sum\limits_{\ell  } {\bar\nu _{\ell}
(x)\gamma _\mu  P_L\ell(x) + \sum\limits_{q,q'} {\bar q(x)\gamma
_\mu  P_LV_{qq'} q'(x)} })+{\rm H.c.},
\end{equation}
where $\ell=e,\mu ,\tau$; $q=u,c,t$; $q'=d,s,b$; $V_{qq'}$ is the CKM matrix,
and the MSSM gluino-quark-squark and neutralino-quark(lepton)-squark(slepton) interactions \cite{hab}
\begin{equation}
\label{glu} {\cal L}_{\tilde{g}}=-\sqrt2g_s\frac{(\lambda_r)^a\,
_b}{2} (\bar{q}_{aL}\tilde{g}^{(r)}\tilde{q}_L^b-
\bar{q}_{aR}\tilde{g}^{(r)}\tilde{q}_R^b)+{\rm H.c.},
\end{equation}
\begin{equation}
\label{chi} {\cal
L}_{\tilde{\chi}}=\sqrt{2}g\sum\limits_{n=1}^{4}(\epsilon_{Ln}
(\psi)\bar{\psi}_L{\tilde{\chi}}^0_n\tilde{\psi}_L+
\epsilon_{Rn}(\psi)\bar{\psi}_R{\tilde{\chi}}^0_n\tilde{\psi}_R)+
{\rm H.c.}
\end{equation}
Here $(\lambda _r )^a\, _b $  are the $3\times 3$ Gell-Mann matrices ($r = 1, \ldots ,8$) with color indices $a,b=1,2,3$; the neutralino coupling constants are defined as
\[
\epsilon_{Ln}(\psi)=-T_3(\psi)N_{n 2}+\tan
{\theta_W}(T_3(\psi)-Q(\psi))N_{n 1},
\epsilon_{Rn}(\psi)=Q(\psi)\tan{\theta_W}N_{n 1},
\]
where $Q(\psi)$ and $T_3(\psi)$ are the electric charge and the third
component of the weak isospin for the quark (lepton) field $\psi$, respectively, and
$N_{nm}$ is the $4\times 4$ neutralino mixing matrix.

The leading order amplitude of the decay (\ref{dec}) is described by 9 diagrams shown in Fig. 1.
\begin{figure}[hbtp]
\hspace{-0.5cm}
\centering
\includegraphics[scale=0.65]{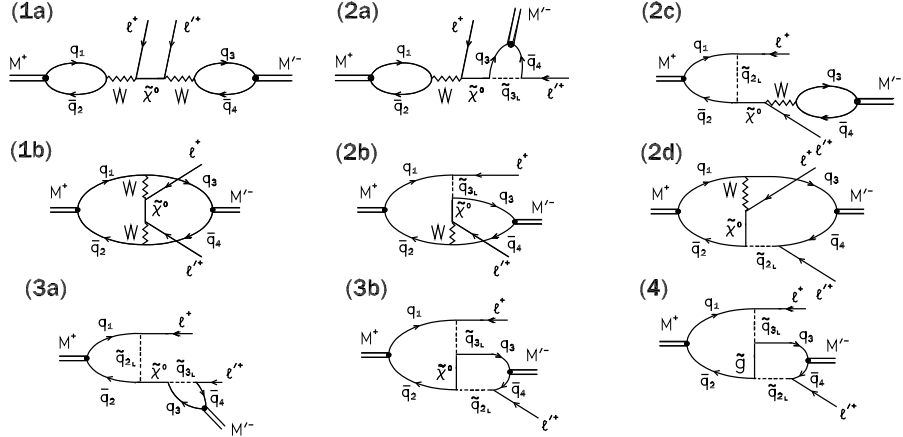}
\caption{Feynman diagrams for the  decay  $
M^{+}\to M'^{-}+ \ell^{+}+ \ell'^{+}$ in the bilinear $\not\!\!R$MSSM. Bold
vertices correspond to Bethe--Salpeter amplitudes for mesons. There are also crossed
diagrams with interchanged lepton lines.}
\end{figure}
The hadronic parts of the decay amplitude are calculated with the use
of a simple model for the Bethe--Salpeter (BS) amplitudes for mesons as
bound states of a quark and an antiquark \cite{est} (see also \cite{abz,abs1}).
In addition, taking into account that the meson mass $m_M  \ll m_W ,m_{SUSY}$, where $m_W$ is the $W$ boson mass and  $m_{SUSY}
\gtrsim 100~\mbox{GeV}$ is the common mass scale of superpartners, we neglect momentum dependence in the propagators of heavy particles (see Fig. 1) and use the effective low-energy current-current interaction. In this approximation the decay amplitude does not depend on the specific form of the BS amplitude and is expressed  through the known decay constants of the initial and final mesons, $f_M $ and $f_M'$.

Finally, for the total width of the decay (\ref{dec}) we have obtained
\begin{eqnarray}
\label{dw}
 &\ds\Gamma_{\ell\ell'}\equiv \Gamma (M^ +   \to M'^ -  \ell ^ +  \ell '^ +  ) = (1 - \frac{1}{2}\delta _{\ell \ell '} )\frac{{g^4 f_M^2 f_{M'}^2 m_M^7 }}{{2^{10} \pi ^3 }}\Phi _{\ell \ell '}^{\rm bi}  \nonumber\\
  &\ds\times \left| {\sum\limits_{n = 1}^4 {\frac{{g^2 }}{{2m_{\tilde \chi _n } }}} } \left[ {\frac{{\kappa_n^{2} }}{{4m_W^4 }}\left( {V_{12} V_{43}  + \frac{{V_{13} V_{42} }}{{N_c }}} \right)} + \frac{{\kappa_n \beta ^{c} }}{{2m_W^2 }}\left( {\frac{{\epsilon _{Ln}^* (q_2 )}}{{m_{\tilde q_{2L} }^2 }}}\left(V_{43}  + \frac{V_{13}}{N_c} \right) \right.\right.\right.\nonumber\\
  &\ds\left.\left.\left.+ \frac{{\epsilon _{Ln}^* (q_3 )}}{{m_{\tilde q_{3L} }^2 }}\left( {V_{12}}  + \frac{V_{42}}{N_c}\right)\right)
  - {\frac{{{\beta ^{c}} ^2 \epsilon _{Ln}^* (q_2 )\epsilon _{Ln}^* (q_3 )}}{{m_{\tilde q_{3L} }^2 m_{\tilde q_{2L} }^2 }}\left(1 + \frac{1}{N_c } \right)}\right]\right.\nonumber\\
  &\ds\left.- \frac{2}{N_c^2}\frac{{g_s^2 {\beta ^{c}}^2}}{{m_{\tilde q_{3L}}^2 m_{\tilde q_{2L} }^2 m_{\tilde g}}}\right|^2.
\end{eqnarray}
Here
\begin{eqnarray}
\label{Phi1}
 &\ds\Phi^{\rm bi} _{\ell \ell '}  = \int\limits_{l_ +  }^{h_ -  } {dz} z^2
\left[ {1 - (h_ +   + h_ -)(2z)^{-1}} \right]^2 \left[ {1 -
(l_ +   + l_ -)(2z)^{-1}} \right] \nonumber\\ &\times \left[
{(h_ + - z)(h_ - - z)(l_ +   - z)(l_ -   - z)}
\right]^{1/2}
\end{eqnarray}
is the reduced phase space integral with $h_ \pm   = (1 \pm m_{M'} /m_M )^2$ and $l_ \pm   = [(m_\ell   \pm m_{\ell '} )/m_M ]^2 $; $N_c  = 3$
is the number of colors.

For the numerical estimates of the branching ratios (BRs), $\mbox{B}_{\ell \ell '} =
\Gamma_{\ell \ell '}/\Gamma _{\rm total}$, we have used the known values for the SM couplings $g$ and $g_s$, meson decay constants, meson and lepton masses \cite{pdg}, and a typical set of supersymmetric parameters \cite{hd}: a) the MSSM parameters: $m_0 = 70~{\rm GeV}$, $\mu = 500~{\rm GeV}$, $M_2 = 200~{\rm GeV}$, $\tan \beta =4$; b) the RPV parameters: $\left| \Lambda  \right| \equiv ( {\sum\nolimits_{i = 1}^3 {\left| {\Lambda _i } \right|^2 } })^{1/2}  = 0.1~{\rm GeV}^2 $ with $10\Lambda _1  = \Lambda _2  = \Lambda _3 $, $\left| \epsilon  \right|^2 \equiv \sum\nolimits_{i = 1}^3 {\left| {\epsilon _i } \right|^2 }  = \left| \Lambda  \right|$ with $\epsilon_1 = \epsilon_2 = \epsilon_3$;  $M_3 =(g_s^2/g^2)M_2$ at the electroweak scale. Using the MSSM mass formulas \cite{hab} with the gluino mass $m_{\tilde g} = M_3$, the masses of squarks and neutralinos and the elements of the neutralino mixing matrix were calculated numerically for the above set of parameters. The results of the calculations with the use of Eq. (\ref{dw}) are shown in the fourth column of Table 1. In the second and third columns of the table, the present direct experimental bounds on the BRs \cite{pdg} and the indirect bounds for the Majorana neutrino mechanism of the rare decays \cite{abs1} are shown, respectively. We see that the BRs for the bilinear RPV mechanism are much smaller than these upper bounds. For comparison, the trilinear RPV mechanism leads to the upper limits on the BRs of order $10^{-23}$ ($10^{-24}$) for $K$ ($D$) rare decays with the use of conservative bounds on the trilinear couplings  $|\lambda '_{ijk} \lambda'_{i'j'k'}|\lesssim 10^{-3}$ \cite{abs2,abs3}. But for more stringent bounds $|\lambda ' \lambda'|\lesssim 5\times 10^{-6}$ \cite{barb,sb}, $\mbox{B}_{\ell \ell^{\prime }}(\mbox{tri}{\not\!R}\mbox{MSSM})\lesssim 10^{-28}$.
\begin{table}[htb!]
\caption{The branching ratios $\mbox{B}_{_{\ell \ell ^{\prime }}}$
for the rare meson decays $M^{+}\to M^{\prime
-}\ell^{+}\ell^{\prime +}.$}
\begin{center}
\begin{tabular}{|c|c|c|c|}
\hline Rare decay &Exp. upper  &Ind. bound
&$\mbox{B}_{\ell \ell^{\prime }}$\\
 &bound on $\mbox{B}_{\ell \ell '}$&on $\mbox{B}_{\ell \ell^{\prime }}$ ($\nu_M$SM)&
$(\mbox{bi}{\not\!R}\mbox{MSSM})$\\ \hline\hline $K^{+}\to \pi
^{-}e^{+}e^{+}$ &$6.4\times 10^{-10}$&$5.9\times 10^{-32}$&$3.6\times
10^{-49}$
\\ \hline

$K^{+}\to \pi ^{-}\mu ^{+}\mu ^{+}$ & $3.0\times 10^{-9}$&$1.1\times
10^{-24}$&$1.0\times 10^{-49}$
\\ \hline

$K^{+}\to \pi ^{-}e^{+}\mu ^{+}$ & $5.0\times 10^{-10}$&$5.1\times
10^{-24}$& $4.2\times 10^{-49}$
\\ \hline\hline
$D^{+}\to K ^{-}e^{+}e^{+}$ & $4.5\times 10^{-6}$&$1.5\times
10^{-31}$&$1.6\times 10^{-48}$
\\ \hline

$D^{+}\to K ^{-}\mu ^{+}\mu ^{+}$ & $1.3\times 10^{-5}$&$8.9\times
10^{-24}$&$1.5\times 10^{-48}$
\\ \hline

$D^{+}\to K ^{-}e^{+}\mu ^{+}$ &$1.3\times 10^{-4}$&$2.1\times
10^{-23}$&$3.1\times 10^{-48}$\\ \hline
\end{tabular}
\end{center}

\end{table}

\end{document}